\documentstyle[12pt,epsf]{article}\textwidth=175mm\textheight=225mm
\makeatletter\@addtoreset{equation}{section}
\makeatother
\begin{document}

\title{\bf Laser cooling of nuclear beams in storage rings}
\author{E.G.~Bessonov $^{\dagger}$, E.V.~Tkalya$^{\dagger \dagger}$\\
\small  $^{\dagger}$ Lebedev Physical Institute RAS, Moscow, Russia\\
\small e-mail: bessonov@x4u.lebedev.ru \\
\small $^{\dagger \dagger}$ Institute of Nuclear Physics, Moscow
State University, Moscow, 119999, Russia\\
\small e-mail:  tkalya@ srd.sinp.msu.ru \\
\date{} } \maketitle

                       \begin{abstract}
Laser cooling of nuclear beams in storage rings is discussed. The main
properties of nuclei with isomeric levels in the energy range $E < 50$
keV are presented and minimal damping times are found when the isomer
half-life in the fully stripped ions is less than some microseconds.
                        \end{abstract}

                     \section{Introduction}

It is under discussion to operate storage rings like HERA, RHIC, LHC
with heavy ions. Because of the heavily increased intrabeam scattering
of heavy ions compared to protons, the achievable luminosity of the
rings is much smaller than desired by the experiments. A way to
increase luminosity could be cooling of the beams. Different methods of
cooling can be suggested. Among them the electron and laser methods of
cooling are the most effective ones.  Because of the high electron
energy necessary for electron cooling of heavy ions the conventional
approach using electrostatic DC sources is not possible. Instead an
electron storage ring could be used. For this ring there are special
requirements like small damping times, emittances and energy spread.
The feasibility of such an electron storage ring was studied for HERA,
and the results with a possible solution are presented in papers
\cite{genter}, \cite{martirosyan}. In this paper the laser cooling of
nuclear beams in storage rings is discussed. The main properties of
nuclei with isomeric levels in the energy range $E < 50$ keV are
considered when the isomer half-life in the fully stripped ion
$\tau^{'\,rad}$ is less than some $\mu$s. There are no any other decay
channels for low energy isomeric nuclei levels, excluding
$\gamma$-radiation, for naked nuclei. That is why, the isomer's half
life in a fully stripped ion coincides with the radiation life time.

      \section{Fundamentals of laser cooling of ions in a storage ring}

As a consequence of Liuville's theorem, the compression of phase space
density for a given ensemble of particles requires a dissipative force.
In the case of both simple and complicated particles such as electrons,
protons, heavier ions moving in the external fields such a force
results from the radiative friction. Friction force appears
through synchrotron/undulator radiation of particles in the fields of
bending magnets, magnetic lenses, undulators or through backward
Compton/Rayleigh scattering of laser photons by simple/complicated
charged particles. Below we will discuss the physics of the backward
Rayleigh scattering of laser photons on relativistic complicated
ions/nuclears and methods of the nuclear beam cooling based on such
scattering \cite {habs} - \cite {kim}.

       \subsection{The Rayleigh scattering of laser photons on
       relativistic ions}

Let a laser beam is directed against and scattered by an ion beam. In
general case the laser beam is not monochromatic one. The central
frequency of the incoming and scattered laser photons in this case
will be written as $\omega _0 ^l$ and $ \omega^s _0$ respectively.
Since we are considering the case near resonance, we have $\omega ^{l}
\simeq \omega^l_0$ and $ \omega^s \simeq \omega^s _0$. Let $\hbar
\omega _{tr}$ be the transition energy in the ion's rest frame between
two electronic (not fully stripped ion) or nuclear (naked ion) states 1
and 2, and $\hbar \omega ^{l}$ and $\hbar\omega^s$ be the corresponding
energies of the incoming laser photons and the scattered photons in the
laboratory frame, respectively. These quantities are related by

        \begin{equation}   
        \hbar \omega _0 ^l = {\hbar \omega _{tr} \over \gamma(1 -\beta
        \cos \psi)}, \hskip 27mm \hbar \omega _0^s = {\hbar \omega
        _{tr}\over \gamma (1 - \beta \cos \theta)},
        \end{equation}
where $\gamma = \varepsilon /Mc^2 = 1/\sqrt{1 - \beta^2}$, $\varepsilon $ is
the ion energy, $M$ its mass, $\beta = v/c$, $v$ the ion velocity, $c$ the
speed of light, $\psi$ and $\theta $ the angles between the initial and
final photon velocities and ion velocity respectively.

In this paper, we restrict to the case $\psi \simeq \pi$, $\beta \simeq
1$, $\gamma \gg 1$, $\theta \ll 1$ in which the above equations become

           \begin{equation} 
           \hbar \omega _0 ^l = {\hbar \omega _{tr}
           \over 2\gamma}, \hskip 25mm \hbar \omega _0^s = {2\gamma
           \hbar \omega _{tr}\over 1 + (\gamma \theta) ^2}.
           \end{equation}

Due to the Lorentz transformations in accordance with (2.2) low-energy
laser photons from the laboratory system are converted to much larger
photon energies in the ion's rest frame. Thus highly charged ions
become accessible to laser cooling.

In the  ion/nuclear rest frame $K ^{'}$ the photo excitation cross
section for photons of the energy $\hbar \omega ^{'\,l}$ is given by
\cite{loudon}- \cite {tkalya}

        \begin{equation}  
        \sigma (\omega ^{'\,l}) = \frac{\pi ^2c^2}{\omega _{tr}^{2}}
        {g_2\over g_1}\Gamma\,g(\omega _{tr},
        \omega ^{'\,l}),
        \end{equation}
where $g _{1,2}=2J _{1,2} + 1$ are the statistical weights of the
states (in the naked ion 1 denotes the nuclear ground state with spin
$J_1$, and 2 denotes the nuclear isomeric state with spin $J_2$),
$\Gamma $ is the natural linewidth for two level system (transition
probability) for the spontaneous photon emission  of not fully stripped
excited ion or nuclear, i.e.  $\Gamma = 1/{\tau^{'\,rad}} $ where
$\tau^{'\,rad} $ is the level decay time in the $K ^{'}$
system\footnote {In nuclear physics another definition is introduced:
$\Gamma  = \ln 2/{\tau^{'\,rad} }$.},

         $$g(\omega _{tr}, \omega^{'\,l}) =
         {1 \over 2\pi} {\Gamma\over (\omega ^{'\,l} - \omega
         _{tr})^2 + {\Gamma ^{2} /4}}
        \hskip 20mm (\int g(\omega ^{'\,l})d\omega ^{'\,l} = 1)$$
is the Lorentzian with normalization condition in the bracket.

The cross-section (2.3) has a maximum $\sigma _{max} = \sigma|
_{\omega ^{'\,l} = \omega _{tr}} = g _2 \lambda _{tr} ^{2}/2\pi g_1$,
where $\lambda _{tr} = 2 \pi c /\omega _{tr}$. It is close to
maximum in the narrow frequency range $\Delta \omega ^{'}/ \omega _0
^{'\,l} = \Delta \omega /\omega _0 ^l \simeq \pm \Gamma/
2\omega _{tr}$.

When the ion beam has an angular spread $\psi _b$ and an energy spread
$\Delta \gamma _b$, then interaction of all ions with the laser beam
occur independently on their angular and energy spreads when the
relative effective bandwidth of the laser spectral line, according to
Eqs (2.1), exceeds the value

             \begin{equation}   
             ({\Delta \omega ^l \over \omega _0 ^l}) _{eff} \, > \,
             {\psi _b^2\over 4} + {\Delta \gamma_b \over
             \gamma},  \end{equation}
where $(\Delta \omega ^l /\omega _0 ^l) _{eff} \simeq max \{\Delta
\omega ^l /\omega _0 ^l, \, \lambda ^l_0 / (1 + \beta) l _{int}\}$ is
the effective laser line width; $(\Delta \omega ^l /\omega _0 ^l)$, the
natural laser line width; $l _{int} = min \{l _{ss}, \, 2l _R, \, l_{wp}
/(1 + \beta)\}$, the effective length of the interaction region, $l
_{ss}$ the length of the straight section of the storage ring; $l _R
= 4 \pi \sigma _{l \,0} ^2 / \lambda ^l_0$ the Rayleygh length of the
laser beam; $l _{wp}$, the length of the laser wevepacket; $\sigma _{l
\,0}$, the transverse dispersion of the laser beam intensity in the
point $s = 0 $ corresponding to the waist of the laser beam; $\lambda
^l_0 = 2\pi c/ \omega _0^l$, the laser wavelength.  The dispersion in
any point $s$ in the longitudinal direction is $ \sigma _l = \sigma _
{l \, 0} \sqrt {1 + s^2/ l _R^2} $.  The terms including the lengths $l
_{ss}$ and $l _R$ take into account the broadening of the laser
spectral line by the finiteness of the interaction region of the ion
and electron beams\footnote {Ions crossing the laser beam will perceive
the monochromatic laser beam as if it has a frequency band $\lambda
^l_0 /(1 + \beta) l _{int}$.}. We suppose that transverse dispersions
of ion beams $\sigma _{x}$ are less than transverse dispersions of
laser beams.

In the ion rest frame, the scattered radiation is spherically
symmetrical and the energies of incident and scattered photons are
about the same. In the laboratory frame, the scattered radiation will
be directed mainly in the ion velocity direction in the narrow interval
of angles $\Delta \theta \sim 1/\gamma$. The energy of the scattered
photons will be Doppler shifted. The maximum and average energy of the
scattered photons are

             $$\hbar \omega_{ max} ^s = (1 + \beta)\gamma \hbar \omega
             _{tr} = (1 + \beta) ^2 \gamma ^2 \hbar \omega _0 ^l,$$

             \begin{equation}   
             \hbar \overline {\omega ^s} = \gamma \hbar \omega _{tr}
             = \hbar \omega ^s _{max}/(1 + \beta).  \end{equation}

In the laboratory frame the laser-ion interaction is represented by the
rate equation, consisting of $n_1 + n _2 =1$ and

          \begin{equation}   
          dn _2/dt = (1 + \beta) [n_1 - (g_1/g_2) n _2] \int \dot n
          _{\omega} \sigma (\omega ^{'l})d \omega - n _2/ \tau.
          \end{equation}
Here $n_1$ ($n _2$) are the occupation probability for ion/nuclear
state $1$ ($2$); factor $(1 + \beta)$ takes into account the motion of
ions toward the photon beam; $\dot n _{\omega} = (1/\hbar \omega ^l)
\partial I ^l/\partial \omega ^l$ is the spectral density of the laser
beam flow; $\tau  = \gamma / \Gamma $, the decay time; in the cross
section $\sigma (\omega ^{'l})$ the laser frequency $\omega ^{'l}$ at
the ion rest frame must be expressed through the laboratory frequency
$\omega ^{l} = \omega ^{'\,l}/ (1 + \beta) \gamma$; the ratio $g_1
/g_2$ represents the symmetry relation for the Einstein coefficients.

Below we assume that the incident laser beam has a uniform spectral
intensity $\partial I ^l/\partial \omega = I ^l/ \Delta \omega ^l$ in
the frequency interval $\Delta \omega ^l /\omega _0 ^l > \Gamma/ \omega
^{tr}$, where $I ^l$ is the total laser intensity. Substituting $n _1 =
1 - n _2$ to the equation (2.6) we will reduce it to

          \begin{equation}   
          dn _2/dt = A - B n _2
          \end{equation}
where $A = [g _2/(g _1 + g_2)](\Gamma /\gamma) D$; $B = (\Gamma
/\gamma)(1 + D)$; $D = I ^l / I_c$ is the saturation parameter; $I _c
= g _1 \omega ^2 _{tr} \hbar \omega ^l _0 \Delta \omega^l/ \pi ^2 c^2
(g _1 + g _2)$ or $I _c [W/cm ^2] = 1.58 \cdot 10^4 [g _1 (\hbar
\omega _{tr}[eV]) ^4/(g _1 + g _2)\gamma ^2] (\Delta \omega ^l/
\omega _0 ^l)$, the saturation intensity; coefficients $A$ and $B$
are proportional to the laser intensity. They differ from zero for
intervals of time when ions go through the interaction regions of the
storage rings.

If the same interaction regions of laser and electron beams are
allocated periodically along the ion trajectory in a storage ring
then the solution of the rate equation in and outside of the
interaction regions is

          $$n _2 (t) = n _2 ^{st} (1 - e ^{-B(t - t _0)}) + n _{2}
          ^{in} e ^{-B(t - t_0)}, \hskip 10 mm (0 \leq t - t_0
          \leq \Delta t _{int}) $$

          \begin{equation}   
          n _2 (t) = n _2 (t _0 + \Delta t _{int}) e ^{t - t _{0}
          \over \tau }, \hskip 10mm (\Delta t_{int} \leq
          t- t_0 \leq T)
          \end{equation}
where $n _2 ^{st} = A/B = [g _2 D/(g _1 + g _2)(1 + D)$, $n _2 ^{in}$
is the initial occupation probability at the entrance of the
interaction region; $t _0$, the time of entrance of the ion in the
interaction region; $\Delta t _{int} = l _{int}/ \beta c$, the
interaction time; $T = C/\beta c N _{int}$; $C$, the circumference of
the storage ring; $N _{int}$, the number of the interaction regions in
the storage ring.

The value $n _2 (t)$ reaches a steady state $n _2 (t) = n _2 (t +
T$) for $t - t _0 \gg \tau $. In this case

          \begin{equation}   
           n _{2} ^{in}|_{t - t _0 \gg \tau} = n _{2 \, max} ^{in} =
           n _2 ^{st} {[1 - e
          ^{-{\Delta t _{int}(1 + D)\over \tau }}] e ^{-{T
          - \Delta t _{int} \over \tau }} \over 1- e ^{-
          {\Delta t _{int}D + T\over \tau }}  }.
          \end {equation}

In particular

          \begin{equation}   
          n _{2 \, max} ^{in}  = \cases {0, \hskip 10mm (\tau
          \ll \Delta t _{int} ) \cr
          n _2 ^{st} {\Delta t _{int} (1 + D) \over T + \Delta t _{int}
          D}, \hskip 10mm (\tau  \gg T)  \cr
          n _2 ^{st}. \hskip 10mm (\tau  \gg T,
          \hskip 4mm \Delta t _{int}D \gg T).}
          \end {equation}

According to (2.8), (2.10) the time dependence of the occupation

          \begin{equation}   
          n _{2}(t) = \cases {n _2 ^{st}, \hskip 10mm (\tau
          \ll \Delta t _{int}, \hskip 10 mm 0 \leq t
          - t_0 \leq \Delta t _{int}) \cr
          0, \hskip 10mm (\tau  \ll \Delta t _{int},
          \hskip 10 mm \Delta t _{int} \leq t - t_0 ) \leq T \cr
          n _2 ^{st} {\Delta t _{int} (1 + D) \over T + \Delta t _{int}
          D}= const., \hskip 10mm (\tau  \gg T)  \cr
          n _2 ^{st}= const, \hskip 10mm (\tau  \gg T
          \hskip 4mm \Delta t _{int}D \gg T).}
          \end {equation}

If $\tau \gg T$, $n _2 = const$ then the quantity $\dot n ^s = n _{2}/
\tau $ can be interpreted as the number of scattered photons per ion
per unit time. It can be presented in the form $\dot n ^s = (1 +
\beta)(I ^l/\hbar \omega ^l)\overline \sigma$, where $\overline \sigma$
is the average cross section of a photon-ion scattering\footnote {This
value agree with the definition of the average cross section determined
by the equation $\overline \sigma = (1/I ^l) \int \sigma_{\omega
^{'\,l}} I _{\omega ^{l}} d \omega ^{l}| _{\Delta \omega ^{l} \gg
\gamma\Gamma } = g _2 \Gamma \lambda _{tr} ^2 /4 g _1 (1 + \beta)
\gamma \Delta \omega ^{l}$ for the case $D = 0$.}

          \begin{equation}   
          \overline \sigma = {\hbar \omega ^l \dot n^s \over (1 +
          \beta) I ^l} = {\pi ^2 c^2 g _2 \Gamma
          \over (1 + \beta) \gamma g _1 (1 + D) \omega _{tr} ^2 \Delta
          \omega ^l}.
          \end {equation}

The average power of the scattered radiation in this case can be
presented in the form

          \begin {equation}   
          \overline {P ^s} = \hbar \overline {\omega _s} \dot n ^s|
          _{\tau  \gg T} = {g _2 \over g _1 + g _2}
          {\hbar \omega _{tr} \over \tau ^{'rad} }
          {N _{int} l _{int} D \over C + N _{int}l _{int} D}.
          \end {equation}

          \subsection {Laser cooling of ion beams in storage rings}

Below we consider a typical scheme for laser cooling of ion beams in
the longitudinal plane based on the resonant interaction of unbunched
ion beam and homogeneous quasi-monochromatic laser beam ($\Gamma /
\omega _{tr} \ll (\Delta \omega ^l / \omega ^l _0) _{eff} \ll \Delta
\gamma _b / \gamma$) overlapping the ion one in the transverse
direction. In this scheme the interaction of the laser and ion beams
start with the most high energy ions at some minimal initial central
frequency of the laser beam. Then the frequency is scanning in the high
frequency direction, and ions of a lower energy begin to interact with
the laser beam and decrease their energy. The scanning of the frequency
is stopped when all ions are gathered at the minimum energy of ions in
the beam.  The rate of the laser frequency scanning must correspond to
the condition $d\varepsilon _r/dt < \overline {P ^s}$, where the
resonance ion energy $\varepsilon _r = Mc^2 \gamma _r$, $\gamma _r = [1
+ (\omega _{tr}/ \omega _0 ^l)^2]/2(\omega _{tr}/\omega _0 ^l)$. The
relative resonance energy of the being cooled ions in the relativistic
case $\gamma _r= {\omega _{tr}/ 2 \omega _{0} ^l}$.

The cooling time in the longitudinal plane and the relative energy
spread of the cooled ion beam are

        \begin{equation}   
        \tau _{\varepsilon} = {2 \sigma _{\varepsilon \,in}\over
        \overline {P ^s}}, \hskip 15mm
        {\sigma _{\varepsilon \,f} \over \varepsilon } \simeq max\left(
        {\hbar \omega _{tr} \over M c^2}, \hskip 2mm {\Gamma
        \over \omega _{tr}}, \hskip 2mm {\Delta \omega ^l
        \over \omega ^l _0} \right).
        \end{equation}
where $\sigma _{\varepsilon \,in}$ is the initial rms deviation of the
ion energy from an equilibrium one, $M$ atomic mass of the ion.  The
minimum damping time, according to (2.13), is determined by

        \begin{equation}   
        \tau _{\varepsilon \,min} = \tau _{\varepsilon}|_{D \gg T/
        \Delta t _{int}} = {g _1 + g _2\over g _1}
        {2\sigma _{\varepsilon \,in}\over \hbar \omega _{tr}}
        \tau ^{' \,rad} = {g _1 + g _2\over g _1}{Mc ^2\over
        \hbar \omega _{0}^l} \delta _{\varepsilon \,in}
        \tau ^{' \,rad}.        \end{equation}
where $\delta _{\varepsilon \,in} = \sigma _{\varepsilon\, in} /Mc ^2
\gamma$. It depends only on the properties of ions and the relative
energy spread of the ion/nuclear beam.  Usually the value $\delta
_{\varepsilon \,in} \simeq (2\div 5) \cdot 10 ^{-4}$.

Laser cooling of not fully stripped non-relativistic ion beams was used
at the TSR ring in Germany and ASTRID in Denmark \cite {shroder} -
\cite{bond}. This is one of possible cooling methods of ion beams in
the longitudinal plane. In other methods the laser frequency can be
constant and ions in the direction of given resonance energy can be
accelerated by eddy electric fields, static longitudinal electric
fields located at some region of the ion orbit or by phase displacement
mechanism.

Bunched ion beams can be cooled in buckets by quasi-monochromatic laser
beam with accelerating fields of radiofrequency cavities \cite{hangst}.
The broadband laser beam with a sharp low frequency edge can be used as
well. In the last case the edge frequency must have such a value that
only ions with energies above the equilibrium one can be excited
\cite{besabstr}.

The considered one-dimensional laser cooling of ion beams is highly
efficient in the longitudinal plane. A special longitudinal-radial
coupling mechanisms such as synchro-betatron resonance \cite {robinson,
sessler,idea2} or dispersion coupling \cite{ oneil, lauer} can be added
for ion cooling both in the longitudinal and in the transverse planes.
In papers \cite{idea2} - \cite{pac} a conventional three-dimensional
radiative ion cooling method is proposed and developed. An efficient
enhanced method of cooling both in longitudinal and transverse planes
was suggested in papers \cite {bkw}, \cite {prstab}. Unfortunately the
damping times in the three-dimensional method of cooling of naked ion
beams are long and the enhanced method of ion cooling in the transverse
plane can be applied only to the not fully stripped ion beams if the
level decay length of ions is shorter then the length of betatron
oscillations in the storage ring.

\section{Nuclei suitable for laser cooling in storage rings}

We have investigated the possibility of the ion beam cooling for
different ions using nuclear transitions. The results of this
investigation are presented in the Table 1. In this Table: $J^{\pi}$
is a nuclear level having spin $J$, and a parity $\pi$; $T _{1/2} ^g$,
the ground state half-life; $E$, an isomeric level energy; $\alpha$, a
conversion coefficient; $T _{1/2} ^{rad}$, a nuclear level half life
for unionized atom. The isomeric level radiative time of a fully
stripped ion in the ion rest frame $\tau ^{'\, rad} = T _{1/2} ^{rad}
/(1+ \alpha)$. The collected nuclei are meet the following
requirements.  They have a ground states half-life $T_{1/2} ^g$ of more
than an year.  They have an isomeric level in the energy range $E < 50$
keV. This level is a first excited level in the nucleus as a rule (if
the isomer is the second excited level, the isomer branching ratio
equals to zero for the decay to the first excited level). The isomer
half-life in the fully stripped (naked) ion $\tau^{'\,rad}$ is less
than some $\mu$s. At this Table we presented also the minimum damping
time $\tau _{\varepsilon \,min}$ of ion beams for $\hbar \omega ^l _0 =
4.0$ eV, $\delta _{\varepsilon \, in} = 2 \cdot 10 ^{-4}$.

The relativistic factor range is $\gamma = 603$ $(^{151}$
Sm($3/2^-,4.821$ keV)) -- 6193 ($^{81}$Kr($9/2^+, 49.55$ keV)) for
the laser photon's energy 4 eV.  For the $^{201}$Hg nucleus
relativistic factor $\gamma = 195$.  There are no experimental values
for the $^{201}$Hg$(1/2^-,1.556 $ keV) isomer half life and conversion
coefficient. We have used estimations obtained in the work
\cite{tkalya-92} for the Table 1.

Below we consider two imaginary examples close to typical ones and one
realistic example for cooling of nuclear beams in the longitudinal
plane.

\underline {Example 1}. Let the ion transition energy $\hbar \omega
_{tr} = 10$ keV ($\lambda _{tr} = 1.241 \cdot 10 ^{-8}$ cm), $\tau
^{'\, rad} = 1 \mu$s, $M = 100 m _u$ ($m _u c ^2 \simeq 931.5$ MeV),
$\gamma = 3000$, $M c ^2 \gamma = 2.79 \cdot 10 ^{14}$ eV, $2\sigma
_{\varepsilon \,in} = 2.79 \cdot 10 ^{10}$ eV, $2 \sigma _x = 0.5$ mm,
$g_2/g _1 = 1$, $C = 27$ km, $N _{int} = 27$, $l _{dec} = 900$ km $ \gg
C/ N _{int}$ \cite {habs}. The laser beam is continuous, $\Delta \omega
^l _0 / \omega ^l _0 = 5 \cdot 10 ^{-8}$, $\hbar \omega ^l _0 = 1.67$
eV ($\lambda _0 ^l = 7446 \AA$). The interaction length $l _{int} = 2 l
_R = 7.5$ m.

In this case the rms transverse size of the laser beam at the waist
$\sigma _{l\,0} = \sqrt {\lambda ^l l _R/4\pi} \simeq 4.4 \cdot 10
^{-2}$ cm, the saturation intensity $I _c \simeq 4.4 \cdot 10 ^5$
W/cm$^2$, and saturation power $\overline P _c = 2\pi \sigma _{l\,0} ^2
I _c = 6.1$ kW. We assume that the optical resonator with conventional
or super-reflection mirrors will permit to decrease the laser power
2$\div$4 orders times \cite {j.chen}, \cite {zhirong-ruth}.  If the
saturation parameter $D = 10$ and the finesse of the resonator $F = 10
^4$, the intraresonator power $\overline P _{L} ^{ir} \simeq 6.1 \cdot
10 ^4$ W and the average laser power $\overline P _{l} \simeq 6.1$ W.
The average power of the scattered radiation $\overline P ^s = 3.49
\cdot 10 ^8$ eV/s, the damping time $\tau _{\varepsilon} = 80$ s$^{-1}$
$\simeq 1.3$ min.

\underline {Example 2}. Let the ion transition energy $\hbar \omega
_{tr} = 2$ keV ($\lambda _{tr} = 6.205 \cdot 10 ^{-8}$ cm), $\tau ^{'\,
rad} = 20 \mu$s, $M = 200 m _u$, $\gamma = 250$, $M c ^2 \gamma = 4.65
\cdot 10 ^{13}$ eV, $2\sigma _{\varepsilon \,in} = 4.65 \cdot 10 ^{9}$
eV, $2 \sigma _x = 0.5$ mm, $g_2/g _1 = 1$, $C = 6.336$ km, $N _{int} =
20$, $l _{dec} = 1167$ km $ \gg C/N _{int}$. The laser beam is
continuous, $\Delta \omega ^l _0 / \omega ^l _0 = 2.5 \cdot 10 ^{-8}$,
$\hbar \omega ^l _0 = 4.0$ eV ($\lambda _0 ^l = 3102 \AA$). The
interaction length $l _{int} = 2 l _{R} = 6.2$ m.

In this case the rms transverse size of the laser beam at the waist
$\sigma _{l\,0} \simeq 2.77 \cdot 10 ^{-2}$ cm, the saturation
intensity $I _c \simeq 5.06 \cdot 10 ^4$ W/cm$^2$, and saturation power
$\overline P _c = 243$ W.  If the saturation parameter $D = 10$ and
finesse $F = 10 ^2$, the intraresonator power $\overline P _{l} ^{ir}
\simeq 2.43 \cdot 10 ^3$ W and the average laser power $\overline P
_{l} \simeq 24.3$ W. The average power of the scattered radiation
$\overline P ^s = 3.31 \cdot 10 ^6$ eV/s, the damping time $\tau
_{\varepsilon} = 23.4$ min.

\underline {Example 3}. Cooling of $^{151} _{62}Sm ^{62+}$ in the HERA
storage ring. Ion transition energy $\hbar \omega _{tr} = 4.821$ keV,
$\tau ^{'\, rad} = 0.322 \mu$s, $M = 151 m _u$, $\gamma = 350$, $M c ^2
\gamma = 4.92 \cdot 10 ^{13}$ eV, $2\sigma _{\varepsilon \,in} = 4.92
\cdot 10 ^{9}$ eV, $2 \sigma _x = 2$ mm,  $g_2/g _1 = 1.5$, $C = 6.336$
km, $N _{int} = 1$, $l _{dec} = 33.81$ km $\gg C$.  The mode-locked
free-electron laser is used. A sequence of laser pulses is built up
inside an optical resonator, $\Delta \omega ^l _0 / \omega ^l _0 = 10
^{-5}$, $\hbar \omega ^l _0 = 6.9$ eV ($\lambda _0 ^l = 1800 \AA$),
$2 \sigma _{l\, 0} = 2$ mm.  The interaction length $l _{int} = 200$ m,
$l _R =300$ m.

In this case the saturation intensity $I _c \simeq 4.2 \cdot 10 ^8$
W/cm$^2$, and saturation power $\overline P _c = 26.4$ MW.  If the
saturation parameter $D = 10 ^{-2}$, a finesse $F = 10 ^2$, the
intraresonator power $\overline P _{l} ^{ir} \simeq 264$ kW then the
average laser power $\overline P _{l} \simeq 2.64$ kW. The average
power of the scattered radiation $\overline P ^s = 2.52 \cdot 10 ^6$
eV/s, the damping time $\tau _{\varepsilon} = 32.9$ min.  The level
decay time $\tau ^{rad} = \gamma \tau ^{' \,rad} = 112.7$ $\mu$s
($\sim$ 5.3 turns) is much less then the period of phase oscillations.
The enhanced method of laser cooling of ion beams can be produced in
the longitudinal plane in a bucket for near the same damping time.

\section {Conclusion}

The considered examples show that naked ion beams can be cooled in the
longitudinal direction for an acceptable damping time $\tau
_{\varepsilon} < 1$ $h$ using lasers of moderate power and high finesse
open resonators in optical/UV regions.

Conventional present day optical lasers ($\hbar \omega _0 ^l = 1 \div
2$ eV) and high finesse optical resonators can be used to cool the ions
$^{201}Hg$, $^{193}$Pt, $^{151}Sm$, $^{171}Tm$, $^{239}Pu$, $^{169}Tm$,
$^{83}Kr$, $^{187}Os$ in the future LHC storage ring\footnote {Although
the wavelength of a laser that has been actually applied for such a
high-finesse Fabry-Perot resonator is limited to 1.064 $\mu$m, the
fabrication technique is sufficiently applicable to shorter wavelengths
\cite {yamane}.}. These ions have transition energies $\hbar \omega
_{tr} < 10$ keV. Almost all ions presented in the Table 1 can be cooled
in LHC  using conventional present day UV lasers ($\hbar \omega _0 ^l
< 7.9$ eV) \cite {ackerman}, Free-Electron Lasers (FEL)
($\hbar \omega _0 ^l < 12$ eV) \cite {neil} - \cite {todd} and UV
resonators  \cite {lambda}.

$^{201}$Hg and $^{193}$Pt  ions can be cooled in the storage ring HERA
for the cooling time $\tau _ {\varepsilon} < 1.0$ h (we suppose $\tau
^{'} _ {\varepsilon}  =20$ $\mu$s for $^{201}$Hg) if conventional
optical lasers ($\hbar \omega _0 ^l \sim 2$ eV) will be used. To cool
in HERA ions with transition energies $\hbar \omega _{tr} \leq 10$ keV
(see above) we can use conventional present day UV lasers ($\hbar
\omega _0 ^l < 7.9$ eV) \cite {ackerman} and powerful free electron
lasers \cite {benson}, \cite {neil2}. At present, multi-kW
free-electron laser (FEL) in UV region ($\hbar \omega _0 ^l \sim 4$ eV,
$\overline P _l = 3$ $kW$) has been planned at Jefferson Lab \cite
{neil2}. Near the same parameters infrared FEL was realized by JAERI
FEL group in Japan. This group has planned the UV laser ($\hbar \omega
_0 ^l \sim 12$) eV as well \cite {minehara}. The capital cost for a
complete facility is estimated to be \$55M in production. The energy of
the electron beams of the superconducting liner accelerators in such
facilities will be $\sim 200$ MeV \cite {todd}.  The width of the
spectral line in free-electron lasers is $\Delta \omega ^l _0 / \omega
^l _0 \sim 10 ^{-3}$. It can be made narrow in the prebunched
(parametric) free-electron lasers using long ($\sim 10 mm$) electron
bunches tuned to a side maximum \cite {aleks2} and a system of coupled
resonators (a mode selection method) \cite {szarmez}, \cite {aleks}. It
follows that conventional lasers and free-electron lasers will be able
to solve the problem of high power optical and UV lasers for cooling of
8 naked ions in HERA and 43 naked ions in LHC in the nearest future.
The power of the present day conventional UV lasers in harder then 10
eV spectral region suitable for cooling another ions is not enough
\cite {benware} - \cite {deroff}.

To cool nuclear beam in the transverse direction we can use
longitudinal-radial coupling of betatron and phase oscillations through
the synchro-betatron resonance.  In this case the cooling time will be
increased $\sim 2$ times.

Notice, that excitation energies of electronic levels of not fully
stripped ions are much lower then nuclear ones. At the same time the
scattering cross-section for electronic transitions is much larger then
for nuclear ones. That is why the effective laser cooling of not fully
stripped ion beams can be realized in storage rings easily. Such beams
can be used for generation of hard X-rays using backward Rayleigh
scattering process \cite {pac}. Unfortunately they can not be used for
experiments with colliding beams because of stripping process of ions
interacting with counterpropagating electron/ion beam.

E.G.Bessonov thanks F.Willeke, S.Levonian for useful discussions and
A.G.Molchanov, I.A.Ar\-tioukov, P.Sergeev for references on UV lasers
and mirrors.

This work was supported partly by the Russian Foundation for Basic
Research, Grant No. 01-02-16199, Grant No 02-02-16209 and Grant No.
00-15-96651 in Support of the Leading Scientific Schools.


\addcontentsline {toc} {section} {\protect\numberline
    {5 \hskip 2mm References}}

\vskip 5mm
{\scriptsize
\begin{table}[hbt] {\normalsize 
Table 1. Damping time ${\bf \tau_{\varepsilon \,min}}$ was calculated
for $\hbar \omega ^l _0 = 4.0$ eV, ${\delta _{\varepsilon} = 2 \cdot
10 ^{-4}}$.               }
\vskip 3mm
\begin{tabular}{|c|c|c|c|c|c|c|c|c|}
\hline
 & \multicolumn{2}{|c|}{\bf Ground State}
  & \multicolumn{4}{|c|}{\bf Isomeric State}
   &
    & \\
\cline{2-7}
{\bf Nucleus}
 & ${\bf J^{\pi}}^{\vphantom{U}}$
  & ${\bf T_{1/2}}$
   & ${\bf J^{\pi}}$
    & ${\bf E}$, keV
     & ${\bf T_{1/2}}$
      & ${\bf \alpha}$
       & ${\bf \tau^{'\,rad}}$
        & ${\bf \tau_{\varepsilon \,min} ^{[sec]}}$
\\ \hline
${^{201}}^{\vphantom{U}}$Hg
 & $3/2^-$
  & stable
   & $1/2^-$
    & 1.556
     & (1--10) ns
      & (2.1--4.7)$\cdot 10^4$
       & (20--500) $\mu$s
        & (3--70)$\cdot 10^2$ \\ \hline
${^{193}}^{\vphantom{U}}$Pt
 & $1/2^-$
  & 50 y
   & $3/2^-$
    & 1.642
     & 9.7 ns
      & 1.2$\cdot 10^4$
       & 116.4 $\mu$s
        & 3.1$\cdot 10^3$ \\ \hline
${^{151}}^{\vphantom{U}}$Sm
 & $5/2^-$
  & 93 y
   & $3/2^-$
    & 4.821
     & 35 ns
      & 920
       & 0.322 $\mu$s
        & 3.8 \\ \hline
${^{171}}^{\vphantom{U}}$Tm
 & $1/2^+$
  & 1.92 y
   & $3/2^+$
    & 5.029
     & 4.77 ns
      & 1408
       & 6.72 $\mu$s
        & 1.6$\cdot 10^2$ \\ \hline
${^{239}}^{\vphantom{U}}$Pu
 & $1/2^+$
  & $2.41\cdot10^3$ y
   & $3/2^+$
    & 7.861
     & 36 ps
      & 5950
       & 0.214 $\mu$s
        & 7.1 \\ \hline
${^{169}}^{\vphantom{U}}$Tm
 & $1/2^+$
  & stable
   & $3/2^+$
    & 8.41
     & 4.08 ns
      & 285
       & 1.17 $\mu$s
        & 27.6  \\ \hline
${^{83}}^{\vphantom{U}}$Kr
 & $7/2^+$
  & stable
   & $9/2^+$
    & 9.396
     & 147 ns
      & 17.09
       & 2.66 $\mu$s
        & 23.1 \\ \hline
${^{187}}^{\vphantom{U}}$Os
 & $1/2^-$
  & stable
   & $3/2^-$
    & 9.746
     & 2.38 ns
      & 264
       & 0.631 $\mu$s
        & 16.5 \\ \hline
${^{133}}^{\vphantom{U}}$Ba
 & $1/2^+$
  & 10.5 y
   & $3/2^+$
    & 12.322
     & 7.0 ns
      & 70.3
       & 0.5 $\mu$s
        & 9.3 \\ \hline
${^{193}}^{\vphantom{U}}$Pt
 & $3/2^-$
  & 50 y
   & $5/2^-$
    & 14.276
     & 2.52 ns
      & 151
       & 0.383 $\mu$s
        & 8.6 \\ \hline
${^{57}}^{\vphantom{U}}$Fe
 & $1/2^-$
  & stable
   & $3/2^-$
    & 14.413
     & 98.3 ns
      & 8.56
       & 0.94 $\mu$s
        & 7.5 \\ \hline
${^{151}}^{\vphantom{U}}$Eu
 & $5/2^+$
  & stable
   & $7/2^+$
    & 21.541
     & 9.6 ns
      & 28
       & 0.278 $\mu$s
        & 4.6 \\ \hline
${^{149}}^{\vphantom{U}}$Sm
 & $7/2^-$
  & stable
   & $5/2^-$
    & 22.507
     & 7.12 ns
      & 29.2
       & 0.215 $\mu$s
        & 2.6 \\ \hline
${^{119}}^{\vphantom{U}}$Sn
 & $1/2^+$
  & stable
   & $3/2^+$
    & 23.871
     & 18.03 ns
      & 5.22
       & 0.112 $\mu$s
        & 1.9 \\ \hline
${^{161}}^{\vphantom{U}}$Dy
 & $5/2^+$
  & stable
   & $5/2^-$
    & 25.6515
     & 29.1 ns
      & 2.35
       & 97.5 ns
        & 1.5 \\ \hline
${^{201}}^{\vphantom{U}}$Hg
 & $3/2^-$
  & stable
   & $5/2^-$
    & 26.269
     & 630 ps
      & 76.7
       & 49 ns
        & 1.1 \\ \hline
${^{227}}^{\vphantom{U}}$Ac
 & $3/2^-$
  & 21.77 y
   & $3/2^+$
    & 27.37
     & 38.3 ns
      & 4.5
       & 0.211 $\mu$s
        & 4.5 \\ \hline
${^{129}}^{\vphantom{U}}$I
 & $7/2^+$
  & $1.57\cdot10^7$ y
   & $5/2^+$
    & 27.8
     & 16.8 ns
      & 5.0
       & 0.101 $\mu$s
        & 1.1 \\ \hline
${^{40}}^{\vphantom{U}}$K
 & $3^-$
  & $1.28\cdot10^9$ y
   & $4^-$
    & 29.83
     & 4.24 ns
      & 0.3
       & 5.52 ns
        & 0.024 \\ \hline
${^{237}}^{\vphantom{U}}$Np
 & $5/2^+$
  & $2.14\cdot10^6$ y
   & $7/2^+$
    & 33.1964
     & 54 ps
      & 185
       & 10.0 ns
        & 0.26 \\ \hline
${^{125}}^{\vphantom{U}}$Te
 & $1/2^+$
  & stable
   & $3/2^+$
    & 35.492
     & 1.48 ns
      & 14
       & 22.2 ns
        & 0.4 \\ \hline
${^{189}}^{\vphantom{U}}$Os
 & $3/2^-$
  & stable
   & $1/2^-$
    & 36.202
     & 0.53 ns
      & 21
       & 11.66 ns
        & 1.5 \\ \hline
${^{121}}^{\vphantom{U}}$Sb
 & $5/2^+$
  & stable
   & $7/2^+$
    & 37.133
     & 3.46 ns
      & 11.11
       & 41.9 ns
        & 0.55 \\ \hline
${^{129}}^{\vphantom{U}}$Xe
 & $1/2^+$
  & stable
   & $3/2^+$
    & 39.568
     & 0.97 ns
      & 12.31
       & 12.9 ns
        & 0.23 \\ \hline
${^{233}}^{\vphantom{U}}$U
 & $5/2^+$
  & $1.592\cdot10^5$ y
   & $7/2^+$
    & 40.35
     & 0.12 ns
      & 350
       & 42.12 ns
        & 1.1 \\ \hline
${^{249}}^{\vphantom{U}}$Bk
 & $7/2^+$
  & 320 d
   & $9/2^+$
    & 41.79
     & 9 ps
      & 126
       & 1.14 ns
        & 0.03 \\ \hline
${^{243}}^{\vphantom{U}}$Am
 & $5/2^-$
  & 7370 y
   & $7/2^-$
    & 42.2
     & 40 ps
      & 160
       & 6.4 ns
        & 0.17 \\ \hline
${^{250}}^{\vphantom{U}}$Cf
 & $0^+$
  & 13.08 y
   & $2^+$
    & 42.722
     & 98 ps
      & 1293
       & 127 ns
        & 8.9 \\ \hline
${^{240}}^{\vphantom{U}}$Pu
 & $0^+$
  & 6550 y
   & $2^+$
    & 42.824
     & 164 ps
      & 928
       & 152 ns
        & 10.2 \\ \hline
${^{246}}^{\vphantom{U}}$Cm
 & $0^+$
  & 4730 y
   & $2^+$
    & 42.851
     & 121 ps
      & 1080
       & 131 ns
        & 9.0 \\ \hline
${^{244}}^{\vphantom{U}}$Cm
 & $0^+$
  & 11.11 y
   & $2^+$
    & 42.965
     & 97 ps
      & 1060
       & 103 ns
        & 7.0 \\ \hline
${^{248}}^{\vphantom{U}}$Cm
 & $0^+$
  & $3.397\cdot10^5$ y
   & $2^+$
    & 43.40
     & 124 ps
      & 1014
       & 126 ns
        & 8.7 \\ \hline
${^{234}}^{\vphantom{U}}$U
 & $0^+$
  & $2.446\cdot10^5$ y
   & $2^+$
    & 43.498
     & 252 ps
      & 724
       & 183 ns
        & 12.0 \\ \hline
${^{238}}^{\vphantom{U}}$Pu
 & $0^+$
  & 87.74 y
   & $2^+$
    & 44.08
     & 177 ps
      & 799
       & 142 ns
        & 9.4 \\ \hline
${^{242}}^{\vphantom{U}}$Pu
 & $0^+$
  & $3.763\cdot10^5$ y
   & $2^+$
    & 44.64
     & 158 ps
      & 759
       & 120 ns
        & 8.1 \\ \hline
${^{238}}^{\vphantom{U}}$U
 & $0^+$
  & $4.468\cdot10^9$ y
   & $2^+$
    & 44.91
     & 203 ps
      & 618
       & 126 ns
        & 8.4 \\ \hline
${^{236}}^{\vphantom{U}}$U
 & $0^+$
  & $2.342\cdot10^7$ y
   & $2^+$
    & 45.242
     & 234 ps
      & 597
       & 140 ns
        & 9.2 \\ \hline
${^{244}}^{\vphantom{U}}$Pu
 & $0^+$
  & $8.26\cdot10^7$ y
   & $2^+$
    & 46
     & 155 ps
      & 642
       & 100 ns
        & 6.8 \\ \hline
${^{235}}^{\vphantom{U}}$U
 & $7/2^-$
  & $703.8\cdot10^6$ y
   & $9/2^-$
    & 46.204
     & 14 ps
      & 65
       & 0.92 ns
        & 0.023 \\ \hline
${^{183}}^{\vphantom{U}}$W
 & $1/2^-$
  & stable
   & $3/2^-$
    & 46.484
     & 188 ps
      & 8.63
       & 1.81 ns
        & 0.046 \\ \hline
${^{232}}^{\vphantom{U}}$U
 & $0^+$
  & 70 y
   & $2^+$
    & 47.572
     & 245 ps
      & 464
       & 114 ns
        & 7.4 \\ \hline
${^{232}}^{\vphantom{U}}$Th
 & $0^+$
  & $1.405\cdot10^{10}$ y
   & $2^+$
    & 49.369
     & 345 ps
      & 332
       & 115 ns
        & 7.5 \\ \hline
${^{81}}^{\vphantom{U}}$Kr
 & $7/2^+$
  & $2.1\cdot10^5$ y
   & $9/2^+$
    & 49.55
     & 3.9 ns
      & 1.3
       & 8.97 ns
        & 0.08 \\ \hline
\end{tabular}
\end{table}
}
\end{document}